\documentclass[conference]{IEEEtran}
\usepackage{flushend}

\usepackage{ifpdf}

\usepackage{cite}

\ifCLASSINFOpdf
  \usepackage[pdftex]{graphicx}
\else
  \usepackage[dvips]{graphicx}
  \DeclareGraphicsExtensions{.eps}
\fi
\usepackage[cmex10]{amsmath}

\usepackage{stfloats}

\usepackage{slashbox}

\usepackage{xcolor}

\begin{document}

%
\title{2D simulation of helical flux compression generator}

\author{\IEEEauthorblockN{S.V.~Anischenko, P.T.~Bogdanovich, A.A.~Gurinovich, A.V.~Oskin}

\IEEEauthorblockA{Institute for Nuclear Problems, Belarusian State University, Minsk, Belarus\\
\& Electrophysical Laboratory,
Minsk, Belarus\\
Email: gur@inp.bsu.by}}


%


\maketitle

\begin{abstract}
Diverse approaches to HFCG's inductance,resistance and armature
expansion calculating are evaluated.
Comparison of simulated and experimentally obtained results is
provided. Validity criteria for different simulation models are
proposed.
Consideration of armature acceleration under the pressure of
detonation products is shown to be beneficial for accuracy of HFCG
simulation. Control of HFCG temperature during simulation enables
detecting critical points of system operation.
\end{abstract}



%
\IEEEpeerreviewmaketitle

\section{Introduction}
Helical flux-compression generators (HFCG) are the useful compact
power sources. Being easily varied by size, shape, coil pitch and
means for initial flux creation, HFCG could be used for various
high-energy and high-current pulsed power applications.
Development of high-performance generators requires accurate
modelling of HFCGs operation by the use of a 2D or even a 3D
approach \cite{1,2,3,4} and consideration of multiple factors
\cite{5,6} affecting HFCG gain, efficiency and output parameters.
Some of theses factors (like 2-pi clocking, crowbar losses,
electrical breakdown etc.) should be apprehended and avoided by
HFCG proper design and manufacturing. Unavoidable effects like
inter-turn proximity effect in the helical coil, diffusion losses
in the vicinity of the contact point between armature and stator,
magnetic field pressure affecting the expanding armature motion,
high-temperature effects and etc. should be taken into account as
thoroughly as possible \cite{7}.

With the goal
to demonstrate
%
{ranges of validity and propose validity criteria for diverse
approaches to HFCG simulation
%
a modular code FCGcalc was developed.}
%
HFCG simulation by this code is presented in the paper.
 FCGcalc
allows selecting multiple options. One- or two-dimensional model
can be used for inductance calculation (''1D'' or ''2D'' options)
\cite{8,1}.
Two approaches are available to describe armature expansion: the
armature expanding part is either considered to be a cone, which
angle is equal to the armature expansion angle (armature velocity)
defined by Gurney equation \cite{Gurney}, or calculated by
equations of a hollow tube motion under the pressure of detonation
products according to \cite{6,FV} (''Gurney'' or ''pressure''
options). Consideration of all nonlinear effects (nonlinear
diffusion, magnetic field pressure) can be turned off to reveal
their influence on FCG operation (''linear'' or ''nonlinear''
options).

Fast high-power HFCGs with moderate parameters loaded by a mostly
inductive load were analyzed with the view of ensuring the
predictive power for performance of the developed HFCG. The set of
specially designed HFCGs were tested in conditions when the most
of factors those could be scarcely considered in simulation were
either avoided or additionally controlled. The present paper
includes simulation approach and results, comparison of simulated
and experimentally obtained results with the very brief
description of HFCG experimental testing.

\section{Basic Physics}
The equation for the performance of ''HFCG+inductive load''
circuit is given with Kirchhof's voltage law by \cite{6,8,9}:

\begin{equation}
    \label{eq1}
    \left( {L_{g} + L_{l}}  \right)\dot {I} + \left( {\frac{{dL_{g}} }{{dt}} +
        R_{g}}  \right)I = 0,
\end{equation}
where ''dot'' means time derivative, $R_{g} $ is the HFCG
resistance including losses of all types, а $L_{g} $ and $L_{l} $
are inductances of HFCG and the load, respectively. Evaluation of
time-dependant $R_{g} $ and $L_{g} $, which are neither dc no ac
resistance and inductance of HFCG, requires consideration of both
mechanical and electrodynamic aspects. Difficulty is caused by the
complicated geometry of HFCG inner volume closed between coil
(which usually is tapered in diameter and has multi-sectional
winding) and expanding armature as well as by the nonlinear
diffusion of magnetic field into HFCG conductors in conditions,
when the current carrying layers on the armature surface are not
fixed.

One-dimensional model for inductance calculation is considered in
\cite{8}, where the HFCG inductance per unit length reads as
follows:

\begin{equation}
\label{eq:1D_inductance} \frac{{dL_{g}} }{{dz}} = \pi \mu _{0}
n^{2}\left( {R_{S}^{2} - R_{A}^{2}} \right) + \frac{{\mu _{0}}
}{{2\pi} }ln\frac{{R_{S}} }{{R_{A}} },
\end{equation}
where $n$ is the winding density, the stator radius $R_{S}$ and
armature radius $R_{A} $ are both functions of z, while $R_{A} $
is also time-dependant.

According to 2D approach \cite{1} the total HFCG inductance is
presented as a sum of axial $L_{z} $ and azimuth $L_{\theta}  $
components due to axial and azimuth currents in HFCG circuit:
$L_{g} = L_{\theta}  + L_{z}$. Contribution due to axial current
can be derived from \cite{6,8}:

\begin{equation}
\frac{{dL_{z}} }{{dz}} = \frac{{\mu _{0}} }{{2\pi}
}ln\frac{{R_{S}} }{{R_{A}}}, \nonumber
\end{equation}
The effective approach to calculation of HFCG inductance
$L_{\theta} $ due to azimuth currents was proposed by \cite{1} and
considered in details in \cite{6}. This approach implies
approximation of HFCG conductors by a set of single-turn
cylindrical sheet perfectly conducting solenoids (rings) of
infinitesimal width $\Delta z \to 0$, each carrying a uniform
current. Rings radii $R_{i} $ are defined as the inner radius of
stator at $z_{i}$ for stator rings and outer radius of expanding
armature at $z_{i}$ for armature rings (expressions describing
expanding armature radius see therein below). The inductance
$L_{\theta}$ can be found from the equality

\begin{equation}
\frac{{1}}{{2}}\sum M_{ij} I_{i} I_{j} = \frac{{L_{\theta}
I^{2}}}{{2}},
\end{equation}
where $I_{i} $ is the current over $i$-th ring, $M_{ii}$ is the
inductance of the $i$-th ring and $M_{ij} $ is the mutual
inductance of two rings with centers distant $x_{ij} $,

\begin{align}
& M_{ii} = \mu _{0} R_{i} \left( {ln\frac{{16R_{i}} }{{\Delta
z}} - 2} \right), \\
& M_{ij} = \mu _{0} \left( \frac{{R_{i}^{2} + R_{j}^{2} +
x_{ij}^{2}
       }}{{\sqrt {\left( {R_{i} + R_{j}}  \right)^{2} + x_{ij}^{2}} } }K\left(
       {\frac{{4R_{i} R_{j}} }{{\left( {R_{i} + R_{j}}  \right)^{2} + x_{ij}^{2}
               }}} \right) \right) \nonumber \\
& -  \mu _{0} \left( \sqrt {\left( {R_{i} + R_{j}}  \right)^{2} +
x_{ij}^{2}} E\left( {\frac{{4R_{i} R_{j}} }{{\left( {R_{i} +
R_{j}} \right)^{2} + x_{ij}^{2} }}} \right) \right), \nonumber
\end{align}
where
\begin{align}
& K = \int^{\pi /2}_0 (1-k^2 \sin^2 \theta)^{-1/2} d \theta \nonumber \\
& \hspace{-2.2 cm} \mbox{ and } \nonumber \\
& E = \int^{\pi /2}_0 (1-k^2 \sin^2 \theta)^{1/2} d \theta
\nonumber
\end{align}
are the complete elliptic integrals of the first and second kind,
$k = \sqrt {{\textstyle{{4R_{i} R_{j}} \over {\left( {R_{i} +
R_{j}} \right)^{2} + x_{ij}^{2}} }}}$
\cite{Maxwell,9}.
Condition $\sum M_{ij} I_{j} = 0$ is valid for perfect conducting
stator and armature. Azimuth current
$I_{i} = n\left( {z_{i}} \right)\Delta zI$
over a stator ring located at axial coordinate $z_{i}$ is fixed by
the circuit total current $I$ and the winding density $n = n\left(
{z} \right)$:
\begin{equation}
n\left( {z = z_{i}}  \right) = \frac{{\sqrt {\left( {2\pi
R_{i} /n_{t}}
           \right)^{2} - \left( {d_{w} + 2\Delta _{w}}  \right)^{2}}} }{{2\pi R_{i}
       \left( {d_{w} + 2\Delta _{w}}  \right)}},
\label{eq2}
\end{equation}
where $R_{i} = R_{S} \left( {z_{i}}  \right)$ and the stator
radius $R_{S}$, the winding wire diameter $d_{w}$, the isolation
thickness $\Delta _{w}$ and the number of winding turns $n_{t}$,
all are the functions of $z$. Azimuth currents over the armature
are conditioned by the stator azimuth currents as follows:
\begin{equation}
I_{A} = - M_{AA}^{ - 1} M_{AS} I_{S}, \nonumber
\end{equation}
where $I_{A}$, is the column vector describing currents over
armature rings, $I_{S} $ is the same for the stator currents,
$M_{AA}$ is the matrix of self and mutual inductances of armature
rings, $M_{AS}$ is the matrix of mutual inductances of armature
and stator rings. The armature expanding part is either considered
to be a cone, which angle is equal to the armature expansion angle
defined by Gurney equation \cite{Gurney}, or calculated by
equations of a hollow tube motion under the pressure of detonation
products according to \cite{FV,6}. In the latter case armature
rings radii at the axial coordinate $z_{i}$ at the instant $t_{i}$
are defined by:

\begin{align}
& R_{i} = \sqrt {r_{i}^{2} + R_{0}^{2} - r_{0}^{2}}  , \\
& \ddot {r}_{i} = 2\pi \frac{r_{i}
            \left( \frac{p_H}{2}
            \left( \frac{r_{0i}}{r_i} \right)^{2k} -
            \frac{R_i - r_i}{r_i} \sigma_D \right)
- \rho_{A} \dot {r}_{i}^{2} \left( \frac{R_i^2 + r_i^2}{R_i} -
2r_{i} \right)}
{m/2 + 2\pi \rho _{A} r_{i} \left( {R_{i} - r_{i}}  \right)}, \nonumber \\
& r_{i} \left( {t_{i}}  \right) = r_{0} ,\,\,\,\,
\frac{d{r}_{i}}{dt} \left( {t_{i}}  \right) = 0, \nonumber
\end{align}
%
where $\sigma_D$ is the dynamic yield strength, $k$ is the
isentrope index for detonation products, $t_i$ is the instant,
when detonation front comes to point $z_{i} $, $p_H$ is the
pressure of detonation products at the Chapman-Jouguet point,
$\rho_A$ is the density of armature material, $m$ is the weight of
explosive per HFCG unit length, $R_0$ and $r_0$ are the initial
armature outer and inner radii, respectively. When the option
''nonlinear'' is selected, the armature deceleration due to
magnetic field pressure $p_{H} = \frac{\mu _{0} n^{2}I^{2}}{2}$ is
included by negative acceleration term as follows

\begin{equation}
\ddot {r}_{Hi} = - \frac{p_{H} R_i \left(t\right)}{\rho _{A}
        hR_{i} \left(0\right)}, \nonumber
\end{equation}
where $h$ is the initial wall thickness of the armature tube. HFCG
resistance $R_{g}$ is calculated by integrating the resistance per
unit length \cite{8}
\begin{align}
\label{eq4}
\frac{{dR_{g}} }{{dz}} = & \frac{{2\pi \eta \left(
{0,t}
       \right)n^{2}}}{{\delta} }\left( {R_{S} + R_{A}}  \right) \\
+ &\frac{{\eta
       \left( {0,t} \right)}}{{2\pi \delta} }\left( {\frac{{1}}{{R_{S}} } +
   \frac{{1}}{{R_{A}} }} \right) \nonumber
\end{align}
over the time-dependant HFCG length. Here

\begin{equation}
    \label{eq3}
    \delta = \left. \frac{H\left( {\xi ,t} \right)}{\partial
                    H\left( {\xi ,t} \right)/\partial \xi} \right|_{\xi = 0}
\end{equation}
is the skin depth determined by nonlinear diffusion of magnetic
field $H\left( {\xi ,t} \right)$ in conductor, $\xi $ is the
distance from the conductor surface deep into, $\eta \left( {\xi
,t} \right)$ is the specific resistance, which is time-dependant
due to conductor heating. For the ''nonlinear'' FCGcalc option
magnetic field diffusion is described by the well know nonlinear
equations (see, e.g. (5.4-5)-(5.4-8) in \cite{7}) with the
boundary conditions

\begin{equation}
\left. {H\left( {\xi ,t} \right)} \right|_{\xi = 0} = \mu _{0}
nI\left( {t} \right). \label{eq6}
\end{equation}
Current $I\left( {t} \right)$ is used to define boundary condition
(\ref{eq6}). The approximate linear approach can be used instead
of exact solution of equations for nonlinear magnetic field
diffusion when the magnetic field induction does not exceed the
critical value $B_{cr} $ (for copper $B_{cr} $=43T) \cite{7}. For
FCGcalc ''linear'' option just this approximate approach is
realized using Eq. (4.1-8) in \cite{7}.

\section{Simulation Code Description}
Simulation input data is organized as follows: stator geometry,
winding parameters, armature tube geometry, explosive parameters,
load and seed source parameters. Stator geometry is set as a
sectioned tapered structure treated as a set of truncated cones
each defined by two face diameters and height. The number of cones
is fixed, but not limited. Winding sections, each defined by the
section length, the number of turns, the number of starts, the
winding wire diameter and isolation thickness, are aligned with
the stator inner surface. Armature tube is defined by its
material, outer diameter and wall thickness. Detonation velocity
and density of explosive define expanding characteristics of the
exploding armature. Depending on the selection of input data the
shape of armature expanding part is considered either to be a
cone, which angle is equal to the armature expansion angle or to
be calculated by equations (\ref{eq2})). HFCG seeding from a
capacitive storage is considered. Seed source (capacitance,
charging voltage, inductance and resistance) and load parameters
(inductance and resistance) are also to be entered. All the
physical constants required for calculation are gathered in a
separate file and can be altered when necessary (they are the
dynamic yield strength, the isentrope index for detonation
products, the pressure of detonation products at the
Chapman-Jouguet point, the density of armature material etc.)
T-zero is set for the instant, when detonation wave enters stator
top end plug (see Fig.\ref{fig:1}). The calculation sequence is as
follows (selected options prescribe what model to be used):

1. Calculation of HFCG geometry functions: $R_{S} \left( {z}
\right)$ and time-dependant armature geometry $R_{A} \left( {z,t}
\right)$;

2. Calculation of inductance $L_{g} \left( {t} \right)$ using
either 1D or 2D model and resistance $R_{g} \left( {t} \right)$
(\ref{eq4});

3. Calculation of HFCG current $I\left( {t} \right)$ by solving
equation (\ref{eq1});

4. Solution of equation for magnetic field diffusion (either
linear or nonlinear);

5. Recalculation of HFCG armature geometry $R_{A} \left( {z,t}
\right)$ and successive recalculation of inductance $L_{g} \left(
{t} \right)$, resistance $R_{g} \left( {t} \right)$ and HFCG
current $I\left( {t} \right)$ as per items 2-3.

Control of HFCG temperature freezes specific resistance growth
with temperature when the temperature of HFCG conducting parts
reaches $T_{HFCG}=1359$K (copper melting temperature).

Output data includes: stator geometry $R_{S} \left( {z} \right)$,
armature geometry $R_{A} \left( {z,t_{cr}} \right)$, crowbar
closing instant, time notches, HFCG ac inductance, $L_{g} \left(
{t} \right)$ and its derivative $\dot {L}_{g} \left( {t} \right)$,
resistance $R_{g} \left( {t} \right)$, HFCG current $I\left( {t}
\right)$ and its derivative $\dot {I}\left( {t} \right)$.

\section{Experiments to be used for comparison}

Reconstruction of HFCG parameters from data acquired in an
experiment mostly looks like interpreting a riddle. HFCG
experimental study commonly provides initial values of HFCG ac
inductance and resistance measured at certain frequency,
parameters of the load and current derivative in the circuit.
Some reference time marks are also usually available enabling
evaluation of average velocities and synchronization of processes
detected by different sensors.

\begin{figure}[htb]
  \centering
  \includegraphics[width=8.5 cm]{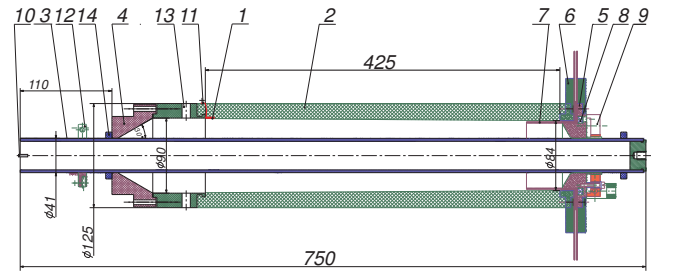}\\
  \caption{HFCG drawing}\label{fig:1}
\end{figure}

The single stage helical flux compression generator with a tapered
coil geometry and winding was specially designed. HFCG stator was
made having biconical inner surface (see Fig.\ref{fig:1}) with the
coil winding 1 aligned with the inner surface of tappered stator
2. Isolated crowbar overlapped the first winding turn which had
inner diameter 90 mm. A copper armature tube 3, which outer/inner
diameter measured 41/35 mm, was centered with the stator by the
top and bottom end plugs 4 and 5, respectively. Sectional top end
plug 4 was made of thermoplastic polymer, while bottom end plug
was made of brass. The conical section of the top end plug 4
served for delicate armature expansion to maximal stator inner
diameter. HFCG output flange 6 and output isolator 7 enabled load
connecting. The bottom end plug had a circular grove, where
Rogowski current derivative monitor 8 was placed. Nut 9 secured
armature tube from axial shift. Seed source output was connected
to buckle 10 and connector 11. Winding length was 425 mm, while
the total stator and armature length measured 570 and 750 mm,
respectively. Overhanging section of armature tube (measured 110
mm) ensured plane detonation wave front coming to the crowbar
location. Optic pins fixed by rings 12 on armature tube at its
both ends enabled measuring the detonation velocity and provided
time marks. Three radial holes 13 in the top end plug were used
for control of armature tube centering accuracy both at the HFCG
assembling stage and during the explosion, when three more optic
pins detected simultaneity of armature expansion to the stator
maximal inner diameter. The same pins enabled measuring the
average armature expansion velocity. Detonation was initiated
end-on. HMX and RDX based explosives with different density values
were used. Load inductance varied from 30 to 150\,nH. Stator
winding included a coaxial end section of either 100~mm or 70~mm
length and several multiple-start sections, which number varied
from 3 to 7 in different tests. HFCG was seeded from the capacitor
bank; the current pulse applied from the seed source was also
recorded.


Experimentally obtained current derivative curve was aligned with
the time marks from optic pins and seed current curve to fix key
points, which were initial data for simulation or could be
compared with simulation results.

\section{Results of Simulation}

\subsection{Comparison of different simulation approaches}

To perceive how difference in the simulation approaches affect the
simulation results, two demonstrative examples are considered. The
stator and armature geometries correspond to those described in
previous section. Two variants of winding are analyzed:

Type 1: 3 sections of 110~mm length each winded by wire of 1.4~mm
diameter with 1, 2 and 4 starts, respectively, and the coaxial end
section of 95~mm length

Type 2: 7 sections of 50~mm length each winded by wire of 1.4~mm
diameter with 1, 2, 4, 8, 16, 32 and 64 starts, respectively, and
the coaxial end section of 75~mm length

Wire isolation thickness is 0.24~mm. Load inductance reads 30~nHn.
Detonation velocity and explosive density are 7~mm/$\mu$s and
1500~kg/m$^3$, respectively.
Initial current in HFCG circuit is defined as current at T-zero
instant. The following figures illustrate when difference between
1D and 2D approached reveals and what is important to choose
armature expansion model.

\subsubsection{Comparison of 1D and 2D models for inductance calculation}

\begin{figure}[htb]
  \centering
  \includegraphics[width=8.5 cm]{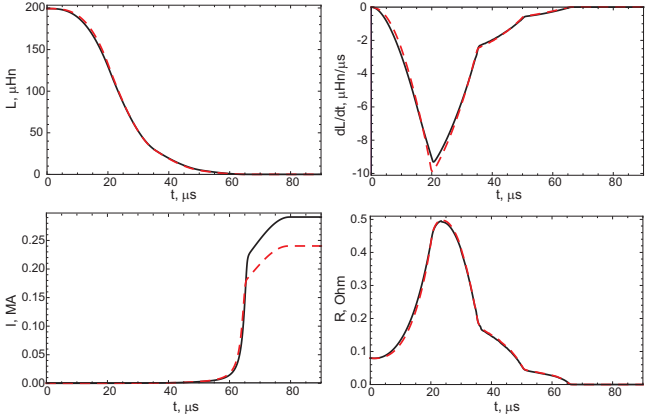}\\
  \caption{Inductance, inductance derivative, resistance and current
  calculated for HFCG with winding of type 1 at 100~A initial current:
  1D model (solid curve) and 2D model  (dashed curve)}\label{fig:3s-100}
\end{figure}

\begin{figure}[htb]
  \centering
  \includegraphics[width=8.5 cm]{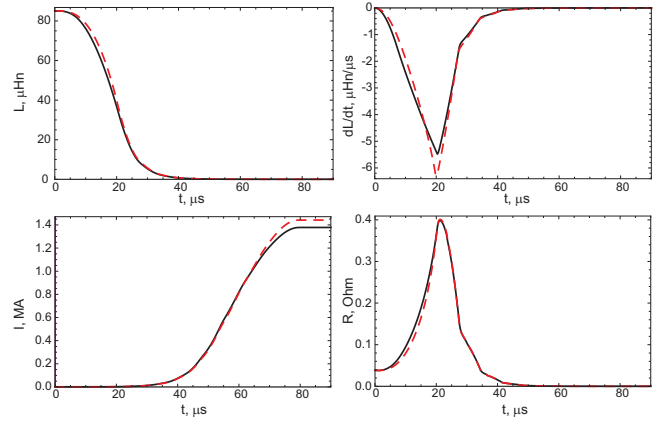}\\
  \caption{Inductance, inductance derivative, resistance and current
  calculated for HFCG with winding of type 2 at 1000~A initial current:
  1D model (solid curve) and 2D model  (dashed curve)}\label{fig:7s-1000}
\end{figure}

Calculation is made for winding of type 1 at 100~A initial current
(Fig.\ref{fig:3s-100}) and for winding of type 2 at 1~kA initial
current (Fig.\ref{fig:7s-1000}), respectively. For both cases no
overheating is expected, as well as the magnetic field induction
value does not exceed $B_{cr}$, so results obtained with
''linear'' and ''nonlinear'' options perfectly coincide. Higher
initial currents give rise to difference in ''linear'' and
''nonlinear'' results. The latter are expected to be valid untill
magnetic field induction exceeds 140~T.

The difference in HFCG current for 1D and 2D models is strongly
pronounced for winding of type 1 because of larger difference in
the inductances of the load and the section previous to the
coaxial end section for the type 1 winding as compared to that of
type 2. For example, for load inductance as high as 300~nH the
relative accuracy of HFCG current predicted by 1D model for
winding of type 1 is approximately the same as for winding of type
2 and load inductance 30 nHn.

\subsubsection{Comparison of approaches to describe armature expansion}

\begin{figure}[htb]
  \centering
  \includegraphics[width=8.5 cm]{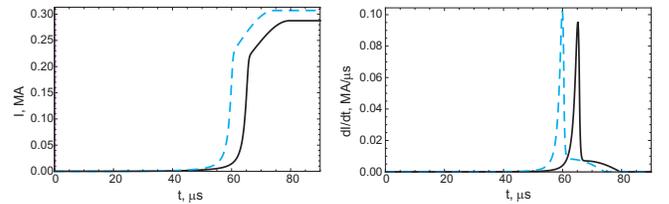}\\
  \caption{Current and current derivative
  calculated for HFCG with winding of type 1 at 100~A initial current:
  ''pressure'' option (solid curve) and ''gurney'' option  (dashed curve)}\label{fig:3s-100a}
\end{figure}

\begin{figure}[htb]
  \centering
  \includegraphics[width=8.5 cm]{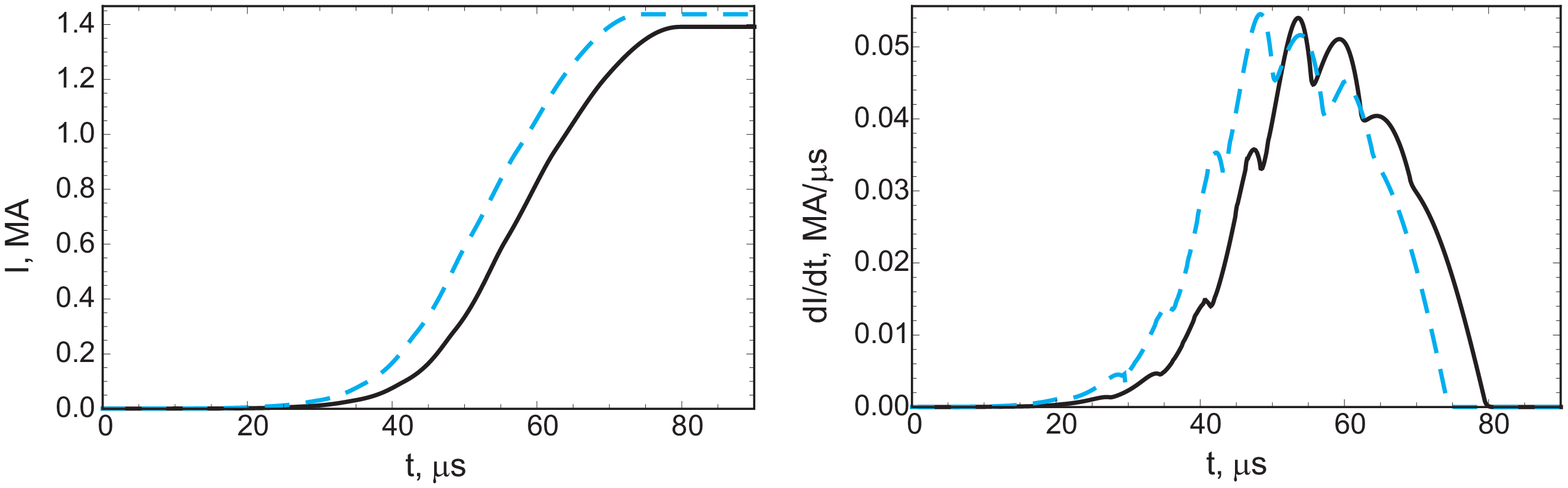}\\
  \caption{Current and current derivative
  calculated for HFCG with winding of type 2 at 1000~A initial current:
  ''pressure'' option (solid curve) and ''gurney'' option  (dashed curve)}\label{fig:7s-1000a}
\end{figure}

Two approaches are used to describe armature expansion: the
armature expanding part is either considered to be a cone, which
angle is equal to the armature expansion angle defined by Gurney
equation \cite{Gurney} (option ''gurney''), or calculated by
equations of a hollow tube motion under the pressure of detonation
products according to \cite{FV,6} (option ''pressure''). The
results of HFCG simulation with these two options mainly differ by
determining crowbar closing instant. In other words they differ by
the time spent for armature expansion to maximal stator inner
diameter.
Almost negligible difference in HFCG current and current
derivative for winding of type 2 (Fig.\ref{fig:7s-1000a}), though
should be compared with the results obtained for type 1 winding
(see Fig.\ref{fig:3s-100a}, where the difference is about 10\%).
When aligning experimental curves with simulated those one should
keep in mind the time difference of this origin.

\subsection{Comparison of simulation and experiment results}

Two experimentally tested HFCGs enable evaluating simulation
reliability.

\noindent \underline{HFCG\#1:}

\begin{itemize}
    \item Winding: 3 sections of 85, 130 and 110~mm length; winded by wire of
1.06~mm diameter with one start, 1.4~mm diameter with one start
and 1.4~mm diameter with three starts, respectively, and the
coaxial end section of 100~mm length;
    \item Load inductance: 30~nHn;
    \item detonation velocity: 7.07~mm/$\mu$s;
    \item explosive density: 1470~kg/m$^3$;
    \item Wire isolation thickness: 0.24~mm.
\end{itemize}

\noindent \underline{HFCG\#2:}
\begin{itemize}
    \item Winding: 5 sections of 75, 65, 80, 65, 70~mm length;
winded by wire of 1.06~mm diameter with one start,1.4~mm diameter
with one start, 1.6~mm diameter with two and four starts and
1.4~mm diameter with eight starts, respectively, and the coaxial
end section of 70~mm length;
    \item Load inductance: 147~nHn;
    \item detonation velocity: 7.9~mm/$\mu$s;
    \item explosive density: 1510~kg/m$^3$;
    \item Wire isolation thickness: 0.24~mm.
\end{itemize}

\begin{figure}[htb]
  \centering
  \includegraphics[width=9 cm]{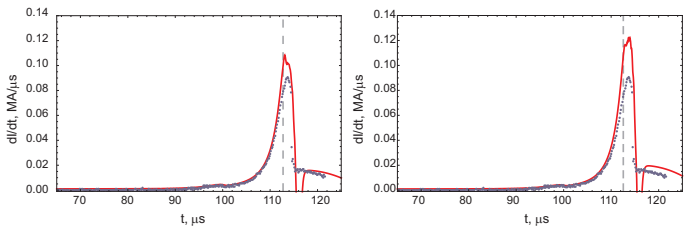}\\
  \caption{Current derivative
  for HFCG\#1 at 700~A initial current and purely inductive load 32nH:
  calculated (solid curve) and experimentally measured  (dotted curve). Left plot is for ''pressure''
  option and right one is for ''gurney'' option}\label{fig:103-32-700}
\end{figure}

\begin{figure}[t]
  \centering
  \includegraphics[width=9 cm]{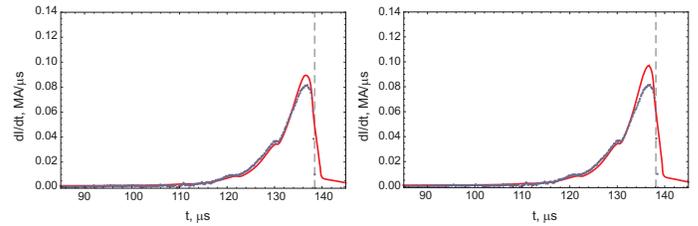}\\
  \caption{Current derivative
  for HFCG\#2 at 800~A initial current and purely inductive load 147nH:
  calculated (solid curve) and experimentally measured  (dotted curve). Left plot is for ''pressure''
  option and right one is for ''gurney'' option.}\label{fig:124-147-800}
\end{figure}

Comparison of experimentally obtained current derivative curves
and those calculated by FCGcalc with ''2D'' and ''nonlinear''
options is presented in Fig.\ref{fig:103-32-700} and
Fig.\ref{fig:124-147-800}. Each figure displays two plots: left
one is for ''pressure'' option and right is calculated for
''gurney'' option. Experimental and simulated curves are aligned
by the mark of last winding section joint with the coaxial end
section. Calculated with ''pressure'' option current derivative
gives better fit for experimental curve for both experiments.
Dashed lines in both figures mark the instant when HFCG conductors
temperature reached copper melting point (overheating mark).
Significant difference in simulated and experimentally obtained
curves arises in the vicinity of overheating mark and apparently
is explained by insufficiency for the extreme conditions. The
latter also explains the negative spike on simulated curve for
HFCG\#1.

\section{Conclusion}

Diverse approaches to HFCG's inductance,resistance and armature
expansion calculating are evaluated and some evaluating
conclusions can be made.

Inductance calculation by 1D model can be used when HFCG
inductance at a contact point approaching the coaxial end section
is comparable with the load inductance. The time difference
provided by different models for armature expansion description
should be kept in mind, when aligning experimental curves with
simulated those. Consideration of armature acceleration under the
pressure of detonation products increases accuracy of HFCG
simulation. Control of HFCG temperature during simulation enables
detecting critical points of system operation.

\end{document}